# Electron motion enhanced high harmonic generation in xenon clusters

Na Li[1], Peng Liu[1,2,★], Ya Bai[1], Peng Peng[1], Ruxin Li[1,2,★] and Zhizhan Xu[1,★]

[1] *State Key Laboratory of High Field Laser Physics, Shanghai Institute of Optics and Fine Mechanics, Chinese Academy of Sciences, Shanghai 201800, China*

[2] *Collaborative Innovation Center of IFSA (CICIFSA), Shanghai Jiao Tong University, Shanghai 200240, China*

**Atomic clusters presents an isolated system that models the bulk materials whose mechanism of HHG remains uncertain, and a promising medium to produce HHG beyond the limited conversion efficiency ($10^{-8} \sim 10^{-5}$) for gaseous atoms. Here we reveal that the oscillation of collective electron motion within clusters develops after the interaction of intense laser fields, and it significantly enhances the harmonic dipole and increases the quantum phase of the harmonics. Experimentally, the phase matching conditions of HHG from nanometer xenon clusters and atoms are distinguished, which confirms the enhanced internal field that was proposed theoretically a decade ago. The separation of HHG from atoms and clusters allows the determination of the amplitude of the HHG for clusters to be 5 orders higher, corresponding to 4 times higher conversion efficiency for atomic response. The finding provides an insight on the HHG mechanism of bulk materials and a means by which an efficient coherent X-ray source can be developed.**

HHG provides a table-top coherent X-ray source that is capable of probing the electronic dynamics[1,2] and seeding the X-FEL laser system to generate fully coherent light at the wavelengths down to "water window"[3]. In the strong field approximation, HHG from isolated atoms or molecules is well explained by the recollision model in which the tunneling ionized electrons are accelerated and recombine with the parent ions to emit high-energy photons[4,5]. Schemes of phase-matching and quasi-phase-matching[6-12] have been proposed to enhance the efficiency of the HHG to the orders of $10^{-5}$. Recently, HHG from bulk materials attracts much interest for the perspective of promoting the conversion efficiency owing to its high density and periodic structures[13-15]. Even though the mechanism is fundamentally different for bulk materials from gases, signature of interband recollisions between electrons and holes has been found in the HHG from zinc oxide[16]. The mechanism remains elusive owing to the lack of information of the electron dynamics in the complex periodic solid structures. Recently, the femtosecond and nanometer visualization of the clusters in intense laser fields confirms the existence of the nanoplasma bound to the cluster Coulomb potential within a few hundreds of femtoseconds before the expansion of the heated clusters[17].

Comparing to the isolated atoms and bulk materials, atomic clusters present an isolated system of high density, offering an ideal model for bridging the mechanisms of HHG from atoms to bulk materials. In clusters, each atom is besieged by the surrounding atoms in a close range through the van der Waals force. Other than the atom itself, the accelerated electrons may recombine with the nearby atoms[18-20], therefore to increase the efficiency and the cut-off energy, but there is still a lack of experimental evidence[21-23]. The cluster-to-itself recollision mechanism considers a ground state wave function that spans in the whole cluster, and the harmonics are generated from the recombining electrons coupling with the delocalized wave function, and the recollision mechanism is supported by the measurement of the group velocity delays of HHG from argon clusters[24,25]. To date, the electron dynamics from ionization to the recombination was treated in strong-field approximation, without consideration of bulk-like environment in the clusters[26].

In this work, we reveal the collective electron motion within clusters dominantly contributes to the HHG. The freed electrons through tunneling ionization are driven into oscillations by the interaction of laser fields and collisions with the cluster boundary. Consequently the resonance of oscillation frequency leads to a transient internal field (TIF) that adds constructively to the laser field. Such a TIF increases the transition dipole greatly, leading to a varied quantum phase and a phase matching condition distinguished from the atoms. In our experiments, the phase-matched HHG from the Xe clusters and atoms are separated and used to identify the enhancement of cluster HHG. By excluding the phase matching factor, the amplitude of the transition dipole for Xe cluster is 5 orders higher than the monomers.

**Results**

**Transient internal field.** As the rare-gas clusters are irradiated by an intense laser field, $\mathbf{E}(t) = E_0 \cos(\omega t)$, in the time scale of tens femtoseconds, electrons are tunneling ionized but remains inside the cluster while the periodic ions (holes) are frozen[17,27]. The electron cloud oscillates around its equilibrium position driven by the external laser fields (Fig. 1). According to the linear theory of collisionless absorption of intense laser field for a cluster system[28,29], when the oscillation of electrons has an amplitude ($\xi$) smaller than the cluster radius, max($\xi$) < $R$, its center position can be solved by the equation of a harmonic oscillator[30,31]. The solution is given by

$$\xi(t) = \text{Re}\left\{\frac{e\mathbf{E}(t)}{m_e(\omega^2 - \omega_M^2 + 2i\gamma\omega)}\right\}, \quad (1)$$

where $\omega$ is the laser frequency, $m_e$ and $e$ are the electron mass and the elementary charge, the Mie frequency is determined by $\omega_M = \sqrt{\frac{4\pi e^2 n_e}{3m_e}}$ which describes the oscillation of the electron density in a spherical cluster, and the damping constant $\gamma$ is given by $\gamma = \gamma_{ei} + \gamma_L$ in which $\gamma_{ei}$ is the frequency of electron-ion collisions that relates to the electrical conductivity of solid materials[32,33], and $\gamma_L = \frac{1}{2R}\sqrt{\frac{8T_e}{\pi m_e}}$ is the electron collisions with the boundary of the cluster, which is inversely proportional to the radius of clusters[29]. The equation (1) describes the motion of electrons in a bulk system that is driven by an external laser fields, reflection of the boundary and the internal plasma environment[31,34]. It is noted that the collision of electrons with the boundary of clusters must be taken into account because of the extreme low rate of individual electron-ion collisions ($\gamma_L \gg \gamma_{ei}$)[26,29,35]. So the solution of equation (1) is in the form of

$$\xi(t) = \xi_0 \cos(\omega t - \phi), \quad (2)$$

with the amplitude term

$$\xi_0 = \frac{eE_0}{m_e\sqrt{(\omega^2-\omega_M^2)^2 + \frac{\omega^2}{R^2}\frac{8T_e}{\pi m_e}}}, \text{ and the phase term}$$

$$\phi = \tan^{-1}[\frac{\omega}{R(\omega^2-\omega_M^2)}\cdot\sqrt{\frac{8T_e}{\pi m_e}}].$$

One can see that the frequency of the electron motion is the same as that of the laser field, while the amplitude and the phase of the motion are dependent on the cluster radius and the Mie frequency.

Such the motion of electrons induces a transient internal field (TIF) in the form[29] of

$$\mathbf{E}_{TIF}(t) = \frac{4\pi}{3}en_e\xi(t), \qquad (3)$$

and the electric field inside the cluster is the sum of external laser field and TIF:

$$\mathbf{E}_{in}(t) = \mathbf{E}_{TIF}(t) + \mathbf{E}(t). \qquad (4)$$

As the electron density increases, resonance occurs at $\omega_M = \omega$ owing to the high density of atoms, where the amplitude of electron motion reaches to the maximum and the phase difference becomes $\pi/2$. An enhanced internal field inside the cluster is then created. However even higher laser intensity will cause the shielding effect of internal field because $\phi = \pi$ for $\omega_M \gg \omega$. For a Xe cluster with the size of $R = 10$ nm under the intense laser field of 20 fs and 800 nm, the displacement of the electron cloud from the equilibrium position in intense laser fields are calculated and shown in Fig. 1. At the moderate laser field of $0.9 \times 10^{14}$ W/cm², the electron oscillation keeps at the phase difference of $\pi/2$ for 4~5 cycles, shown in Fig. 2a. As one can see, the TIF adds to the driving laser field constructively and results in the enhanced internal field $\mathbf{E}_{in}(t)$, shown in Fig. 2c. While in Fig. 2b the phase difference of $\pi/2$ lasts only ~1 cycle at the higher laser intensity of $3.0 \times 10^{14}$ W/cm², and the shielding effect of internal field kicks in as the phase retardation becomes close to $\pi$, shown in Fig. 2d.

The enhanced absorption of intense laser fields for clusters is consistent with the scenario that has been proposed using the method of time-dependent density function theory (TDDFT)[26,36]. The calculation reveals the enhanced local field which explains the dynamic ionization ignition that have been investigated experimentally[37-40].

**High harmonic generation in clusters.** The enhanced field can not only amplify the harmonic dipole, but also vary the quantum trajectory of electronic wavepacket for HHG. Our calculation (see Supplementary Fig. 1) indicates 5 orders enlargement for the harmonic dipole of a single Xe cluster compared to a single atom.

The phase match condition for HHG demands the minimum mismatch factor[41] $\Delta k$

$$\Delta k = \delta n_{disp} \times \frac{q\omega}{c} + \delta n_{plasma} \times \frac{q\omega}{c} + \Delta k_{geom} + \Delta k_{quantum}, \quad (5)$$

in which the first three macroscopic terms represent the dispersion, plasma dispersion, geometric phase mismatch respectively, which are the same for the coexisting clusters and monomers under the supersonic expansion to form the clusters, and the last term $\Delta k_{quantum}$ is determined by the intrinsic phase of the re-colliding electrons

$$\Delta k_{quantum} = \frac{d[q\omega t_f - S(\mathbf{p}_{st}, t_i, t_f)]}{dI} \times \nabla I, \quad (6)$$

where the $S$ is the quasi-classical action for the electron born at $t_i$ with a canonical momentum $\mathbf{p}$ and returning to the origin at $t_f$, namely the quantum phase[5,42]. For clusters the quasi-classical action $S_c$ can be written as

$$S_c(\mathbf{p}_{st}, t_i, t_f) = \int_{t_i}^{t_f} dt'' \left( \frac{[\mathbf{p}_{st} - \mathbf{A}_c(t'')]^2}{2} + I_P \right), \quad (7)$$

where $\mathbf{A_c}$ is the tailored vector potential defined by $\mathbf{E}(t) + \mathbf{E}_{TIF}(t) = -\frac{\partial \mathbf{A}_c(t)}{\partial t}$. Notably the additional TIF term induces a modified quantum phase different from the monomers, which results in a varied phase matching conditions for clusters.

**Experimental observations and discussion.** Experimentally (see Methods), supersonic expansion Xe atoms collide to form clusters within a range of 0.4 mm out of the nozzle[43,44], after which the supersonic beam goes to a non-collision zone due to the jet expansion in vacuum. The collisionless region in the jet downstream provides a condition under which the number densities of the atom and clusters decrease at the same rate[45] while all other parameters are identical. We record the harmonic spectra as a function of the downstream position of the supersonic jet (*y*-scan), as shown in Fig. 3. The averaged harmonics intensity (of orders H15 ~ H19) as a function of the *y* position is shown in Fig. 4a. As the stagnation pressure is below 0.4 MPa, the intensity of HHG presents mainly a shifting peak (labelled A) as the pressure increases. The peak is considered as the result of the phase matching condition as the Xe atom density decrease and the medium length increase during the jet expansion. It is interesting to note that as the pressure increases to ≥ 0.5 MPa, the second intensity peak (B) appears, and the separation of the two peaks becomes larger following the higher pressure.

The yield of HHG from the Xe atoms is expressed as[46,47]

$$I_q \propto n_0^2 |q\omega|^2 |d(q\omega)|^2 \left[\frac{1+e^{-2\alpha_s L} - 2e^{-\alpha_s L}\cos(\Delta k L)}{\alpha_s^2 + \Delta k^2}\right], \quad (8)$$

where $n_0$, $d(q\omega)$, $\alpha_s$ and $\Delta k$ represent the atomic number density, the *q*th-order harmonic strength of a single atom, the field absorption coefficient and the phase mismatch parameter, respectively. The equation (8) can be considered as the product of three parts: one is the single-atom response

$$I_{single} \propto |q\omega|^2 |d(q\omega)|^2, \quad (9)$$

the second is $n_0^2$ and the last part is the phase matching coefficient

$$k_{PM} = \left[\frac{1+e^{-2\alpha_s L} - 2e^{-\alpha_s L}\cos(\Delta k L)}{\alpha_s^2 + \Delta k^2}\right]. \quad (10)$$

We calculate the coefficient $k_{PM}$ ~*y* as the backing pressure increases, which shows a shifting peak that is consistent with the peaks A (Fig. 4). But there is no secondary peak appears for the phase matching condition of the atoms. This indicates that the peak B may be attributed to the phase matching of clusters formed in the supersonic jet, within which an estimated 0.00021% of the particles are clusters[24,43].

There are two observations that support the assignment of peaks B to the phase match condition of clusters. Firstly, by considering the cluster formation within the supersonic jet we calculate the $k_{PM}$ ~*y* following the increasing stagnation pressure. The result shows two series of shifting phase matching position that can completely match the peaks A and B (Figure 5a). Secondly, we plot the harmonic intensity of the peaks A and B as the function of average atomic density (Fig. 5b). One can see that the atomic density dependence of the peak A and B indicates an order of 1.8 and 3.1, respectively. It is consistent with the previous studies that indicates a higher-order pressure dependence of HHG for clusters than the monomers[22] (the average atomic density is proportional to the pressure). The fitting value of 3.1 supports our assignment of peak B the harmonics emission from the clusters.

With the separation of the phase match conditions of clusters and atoms, the ratio of their single-particle response $I_{single}$ can be retrieved from equation (9). Take the result of back pressure 1.0 MPa as a sample, the ratio of $I_{single}$ for clusters and atoms is obtained from the HHG intensity of peak B and A and their number densities, $I_{single}$(cluster): $I_{single}$(atom) = $3.7 \times 10^{10}$. For the averaged atomic response in clusters, $I_{atom}$(cluster): $I_{atom}$(atom) = 4.1 from the estimated number of Xe atoms N = $9.5 \times 10^4$ in cluster. The results is similar to that in ref. 22, in which the atomic yields of HHG from clusters was estimated to exceed the monomers by a factor of 3~5. Our result implies that a Xe cluster (N = $9.5 \times 10^4$) can produce HHG with the energy of 10 orders larger than an atom, which comes from coherent superposition of all atomic HHG within clusters. Here the density of clusters is low comparing to atoms in a jet. Given a way to improve the density of the clusters, the efficiency of HHG can be increased significantly.

Our experimental result (Fig. 4a) shows that the separation of the phase matching condition of clusters and atoms appears only after the cluster grows to a certain size, estimated N = 10000. The size dependence indicates the transition from the atomic recollision mechanism to the bulk mechanism. For clusters in a smaller size (~ 1 nm), the ionization rate is low owing to the lower ionization potential[48], and the collective motion of electrons cannot be established effectively[26,29], so the induced internal field is negligible. The recollision model considering the delocalized initial-state wave function shows the characteristics assembling the gaseous atoms. As the clusters grows to a larger size ($R$ ~10 nm), the ionization rate increases and the collective motion of electrons starts to play a dominant role, so the internal field is enhanced. Because the resonance width of plasma in clusters is larger than that in solid material, the enhanced internal field can last longer time under the time scales of interaction with laser pulse, unlike the bulk medium. As the clusters grows to be the bulk size ($R$ >100 nm), the electron collisions with the boundary of the cluster can be ignored, so the $\gamma_{ei}$ dominates, and the HHG goes into the regime of solid.

It is known that the quantum phase mismatch is linearly dependent on the divergence of laser field

$$\Delta k_{quantum} = a_c \times \nabla I, \qquad (11)$$

where $a_c$ can be calculated from the phase of HHG as a function of laser intensity (see Supplementary Fig. 2). The result indicates that the $a_c$ for Xe clusters is 6 times larger than that of atoms. This may well explain the previous finding that the cluster HHG appears as an off-axis emission in the cut-off region[24,42]. As the TIF adding to the intense laser field constructively, it is expected that the cut-off extends to the higher energy region. However, the extension of the cut-off energy has not been observed in our experiment. Notably, H. Park *et al.*[25] did not observe the extension either. We postulate that the emission of the cutoff is originally weak and in a large divergence angle, which make it difficult to detect for our detection system with a long length from the HHG source to the detector. For the efficient HHG from clusters one needs to develop effective phase matching conditions. The established schemes of quasi - phase - matching (QPM)[9], flat-top driving laser beam[10,11] and slowly varying long-length focused driving field[12] may be adopted.

**Methods**

The experiments (shown in Fig. 3a) were carried out using the 0.7 mJ, 35 fs, 800 nm pulses at 1 kHz, delivered by a Ti: sapphire amplifier laser system (Coherent, Inc. Elite-HP-USX). The pump pulses were focused at 2 mm in front of the pulsed xenon (Xe) supersonic jet with a 500 μm diameter orifice. The nozzle is equipped with a cone of 30 ° half-angle and the backing pressure is tunable from 0.2 MPa to 1.1 MPa to ensure clusters are formed in varied sizes (N = $2 \times 10^3$ ~ $1.2 \times 10^5$), characterized by the Hagena empirical parameter[49]. The supersonic jet is mounted on a mortorized stage and its position is adjustable with the minimum step size of 0.05 mm in the vacuum along the directions of laser propagation ($z$-axis) and jet expansion ($y$-axis). The spectra of HHG were recorded using a home-build soft x-ray spectrometer consisting of a spherical gold mirror, a cylindrical gold mirror, a flat field grating, and a soft X-ray charge-coupled device (CCD) camera (Andor DO440-BN). A typical harmonics spectrum including the emission of the 13th ~ 19th harmonics is shown in Fig. 3b.

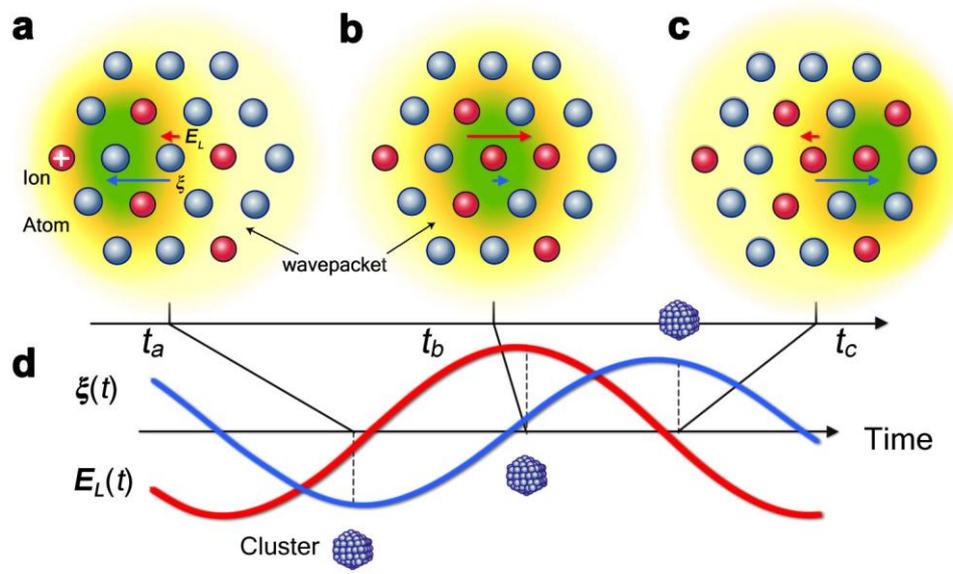

**Figure 1 | Schematic diagram of a time-dependent collective electron motion in a cluster driven by the external intense laser fields.** (**a**), (**b**) and (**c**) illustrate the electronic distributions at the different time of laser pulses, $t_a$, $t_b$ and $t_c$ in (**d**), respectively; (**d**) The laser field (red line) and the center position of electron cloud (blue line) in the cluster.

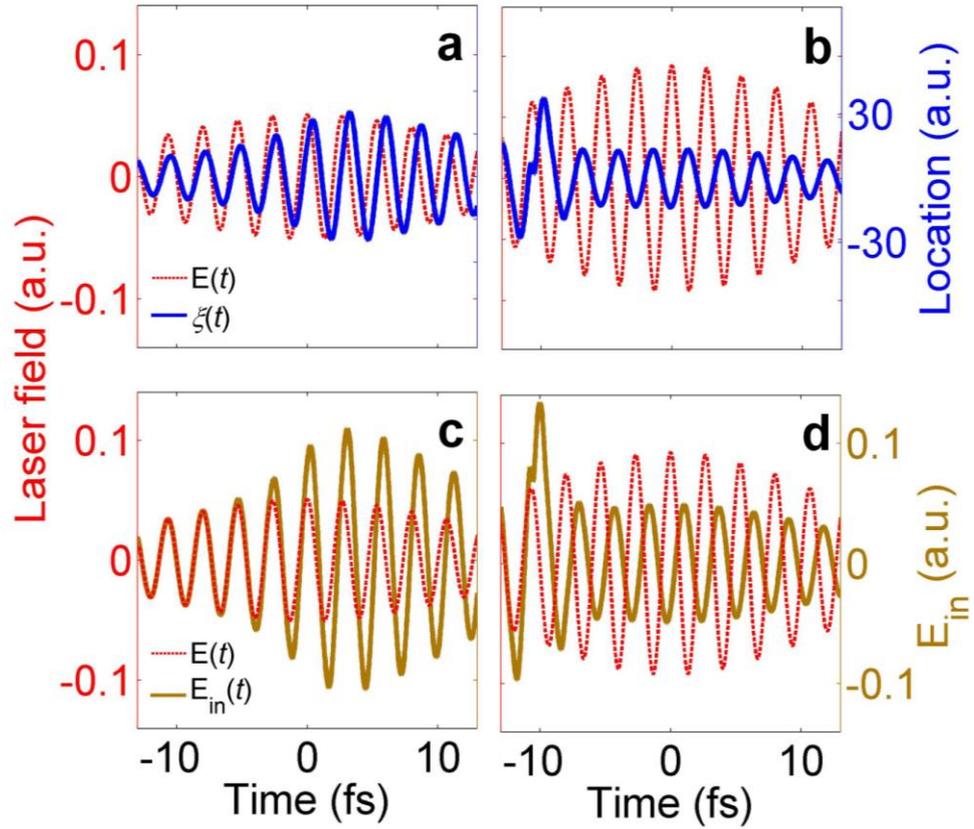

**Figure 2 | Simulated displacement of electrons and internal field for different laser parameters.** For a Xe cluster ($R$ = 10 nm) under a laser field of 20 fs and 800 nm, the displacement of electrons $\xi(t)$ (blue lines) at the laser intensity of (**a**) $0.9 \times 10^{14}$ W/cm$^2$ and (**b**) $3.0 \times 10^{14}$ W/cm$^2$, and the internal field $\mathbf{E_{in}}(t)$ at the laser intensity of (**c**) $0.9 \times 10^{14}$ W/cm$^2$ and (**d**) $3.0 \times 10^{14}$ W/cm$^2$.

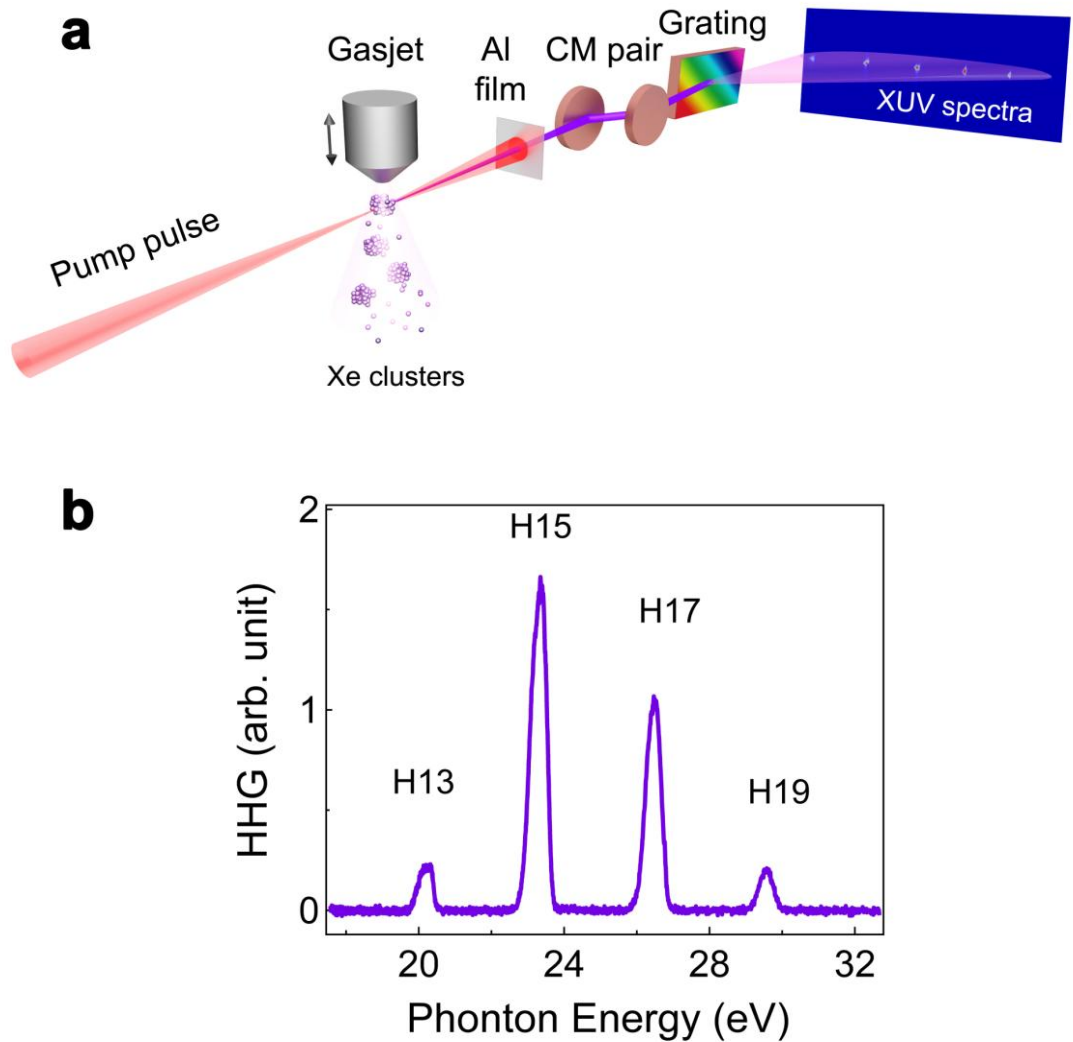

**Figure 3 | Experimental design and HHG spectra. (a)** Experimental setup for high-order harmonic generation from gaseous atoms and clusters; (**b**) Harmonic spectrum of Xe atoms and clusters at back-pressure of 0.4 MPa.

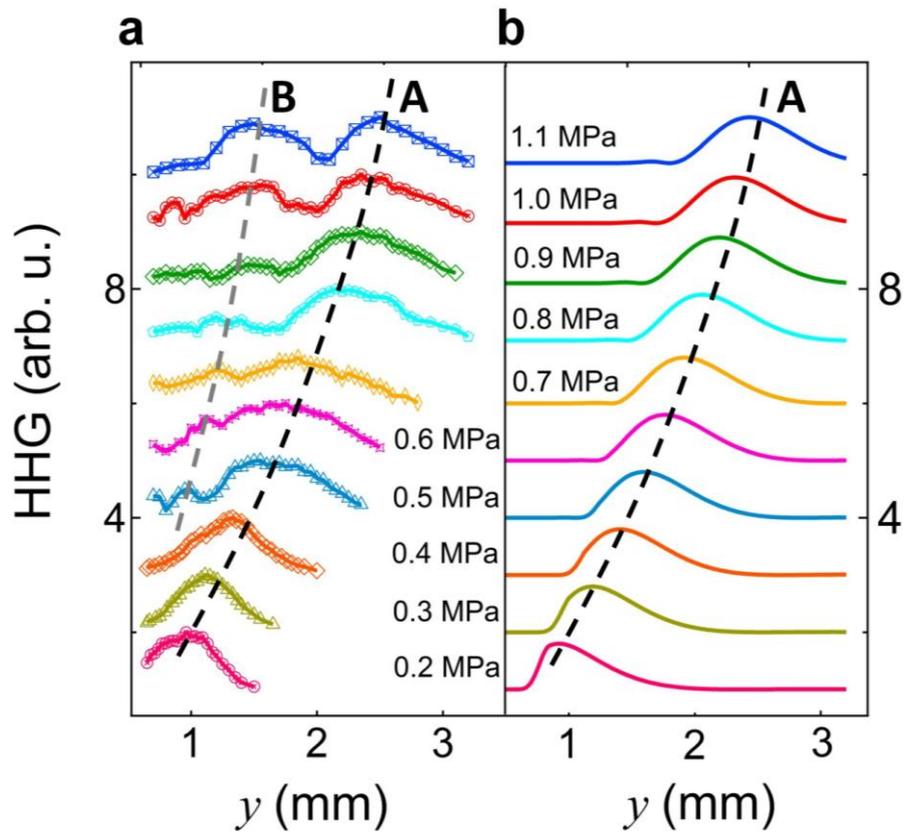

**Figure 4 | Measured HHG and simulated $k_{PM}$ of atoms.** (**a**) Normalized intensity of HHG measured from Xe jet as a function of the downstream position (*y*-scan); and (**b**) simulated phase matching coefficient $k_{PM}$ of the 15th-order harmonics considering only the Xe atoms.

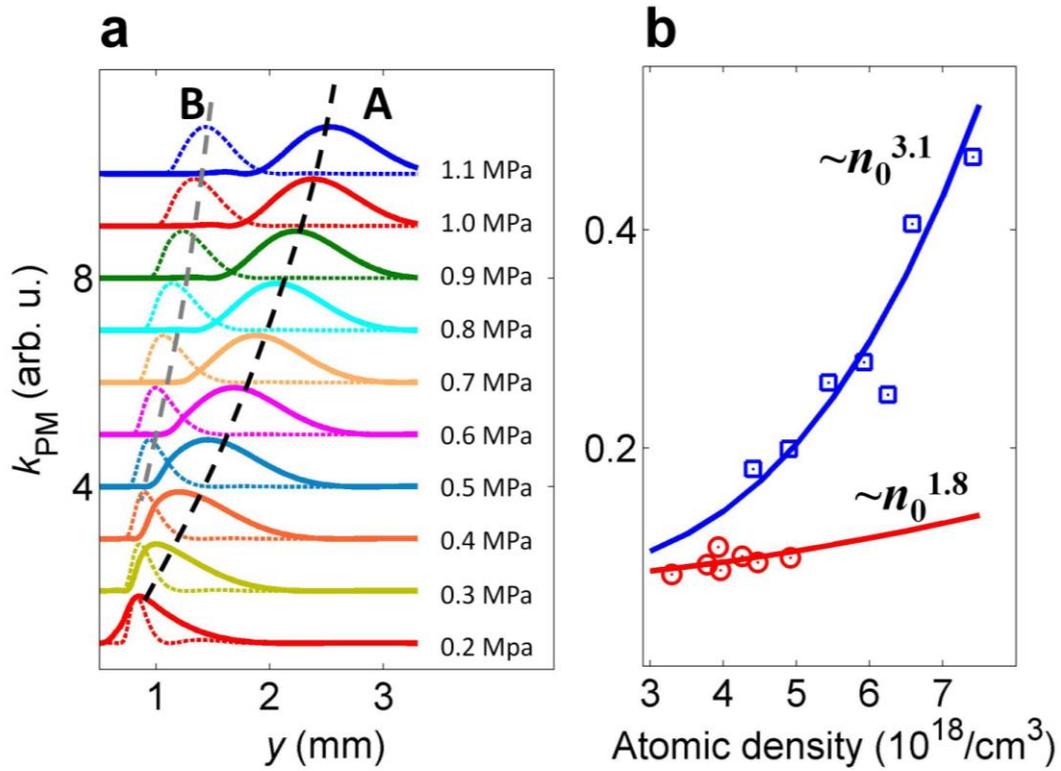

**Figure 5 | Simulated $k_{PM}$ of the mixture and retrieved HHG yield.** (**a**) The simulated phase matching coefficient $k_{PM}$ as a function of the position ($y$-scan) for a mixture of atoms (solid line) and clusters (dotted line); and (**b**) the retrieved HHG yield (H15) at different atomic density.